\documentclass[%
 reprint,
groupedaddress,
 amsmath,amssymb,
 aps,
]{revtex4-1}
\usepackage{braket}
\usepackage{graphicx}
\usepackage{dcolumn}
\usepackage{bm,color}
\usepackage[normalem]{ulem}

\usepackage{amsmath}

\newcommand{\red}[1]{\textcolor{black}{#1}}

\newcommand{\blue}[1]{\textcolor{black}{#1}}
\begin{document}

\title{Emergence and Ordering of Polygonal Breathers in Polariton Condensates}

\author{Samuel N. Alperin${}^1$ and Natalia G. Berloff${}^{1,2}$}
\email[correspondence address: ]{N.G.Berloff@damtp.cam.ac.uk}
\affiliation{${}^1$Department of Applied Mathematics and Theoretical Physics, University of Cambridge, Cambridge CB3 0WA United Kingdom,\\ ${}^2$Skolkovo Institute of Science and Technology Novaya St.,100, Skolkovo 143025, Russian Federation}

\begin{abstract}
We show that the simultaneous driving of a polariton condensate with both nonresonant and $n^{th}$ order resonant pump frequencies allows for a generic mechanism of breather formation\red{; from this we construct for the second order resonance a family of exotic breathers with nontrivial discrete order of rotational symmetry}. Finally, we demonstrate the spontaneous emergence of both crystalline and glassy orderings of lattices of polygonal breathers, depending on the degree of polygonal excitations at the lattice sites.

\end{abstract}

\maketitle

\textit{Introduction- }
The basic nonlinear excitations of Bose-Einstein condensates (BECs) are of significant fundamental interest, and have been studied in detail for decades \cite{perez1997dynamics,matthews1999vortices,madison2000vortex,denschlag2000generating,anderson2001watching,dutton2001observation,penckwitt2002nucleation,morsch2006dynamics,henn2009emergence,navon2016emergence}. However, relatively little is understood about their breather solutions \cite{saint2019dynamical}. In atomic BECs, there are significant fundamental restrictions on the formation and stability of breathers, due to their intrinsic tendency towards thermodynamic equilibrium. Solutions with sustained density oscillations can be constructed by the superposition of the ground state with one of the eigenstates of the Bogoliubov excitations, however, these simple periodic solutions are only persistent in the limit of zero amplitude so as to avoid damping via nonlinear spectral broadening, and lose periodicity as modes are mixed \cite{saint2019dynamical}. Other simple breathing solutions have been constructed with the help of explicit periodicity of the potential in space \cite{trombettoni2001discrete} or of the interaction term in time \cite{matuszewski2005fully}. In a non-periodic system, Pitaevskii and Rosch showed that in two spatial dimensions the nonlinear Schrodinger equation under harmonic trapping admits solutions in which the potential energy oscillates without damping, due explicitly to the SO(2,1) dynamical symmetry of that system \cite{pitaevskii1997breathing,chevy2002transverse}. This latter phenomenon has recently been extended, with it being experimentally and theoretically shown that the SO(2,1) symmetric system also allows particular solutions which are periodic in the wavefunction evolution \cite{saint2019dynamical,lv2020s}.

Due to its inherent nonequilibriation and strong nonlinearity, BEC of exciton-polariton (polariton) quasiparticles has quickly established itself as a central object of study in nonequilibrium quantum dynamics \cite{deng2010exciton,keeling2011exciton,byrnes2014exciton}. By their nature these condensates are not required to conserve particle number, their populations instead ebbing and flowing as a part of their dynamics. \red{Even in the steady state } the constant dissipation and \red{excitation} of quasiparticles makes for a quantum fluidic system in which \red{stationary flows connect spatial regions from where particles are created to where they dissipate}; states of constant density can exist, but represent persistent flows of unchanging geometry. This nonconservative quantum hydrodynamics, while much more complicated than the well developed theory of conservative quantum hydrodynamics, presents a vast landscape of nonequilibrium pattern forming behavior in the setting of a macroscopic quantum fluid \cite{pismen2006patterns}. The polariton BEC bridges the gap in behavior between two of the best studied extended nonlinear systems in physics: the atomic BEC and the nonlinear optical resonator \cite{keeling2011exciton}.

To form and sustain a polariton condensate, the cavity in which it lives must be forced optically. These input photons may be either resonant or nonresonant with the natural frequency of the cavity. Thus, understanding the fundamental repercussions of the forcing type is among the most fundamental problems in the rapidly growing field of polaritonics.

In this Letter, we show that in polariton BECs a generic mechanism of breather formation arises from the combination of nonresonant and resonant forcing, due to the competition between the distinct symmetries associated with these forcing types. Focusing on the special case of $2^{nd} - $ order resonance (maximizing the tension between forcing symmetries), we both explain the recent discovery of breathing ring solitons \cite{alperin2020formation}, and construct a new family of breathers, in which rotational symmetry is spontaneously broken in lieu of polygonal (dihedral) spatial symmetry, with the degree of the resulting polygonal breathers set by the spatial extent of resonant pumping. We show that lattices of the emergent polygonal breathers can spontaneously adopt crystalline or glassy orderings of their orientation, in spinless analogues of ferromagnetic and spin glass  orderings.

\textit{Generic Breathing Mechanism-} A prototypical example of the condensed Bose gas driven far from equilibrium, the dynamics of the polariton condensate can be well described by a generalized complex Ginzburg-Landau equation  (cGLE). This can be written nondimensionally for the condensate wavefunction $\psi({\bf r},t)$ and for the reservoir of uncondensed particles $N_R({\bf r},t)$, as \cite{kalinin2018simulating,keeling2008spontaneous, wouters2007excitations, carusotto2013quantum,keeling2011exciton}
\begin{eqnarray}
  \label{GPE}
 i \partial_t\psi &= &
 -(1-i\eta N_R) \nabla^2 \psi
    + |\psi|^2 \psi+g N_R \psi \\ \nonumber &+& i(N_{R}-\gamma)\psi+i \bar{P}\psi^{*(n-1)} \label{Psi}\\
 \partial_t N_R&=&P-(1+b|\psi|^2)N_R, \label{NR}
 \end{eqnarray}
 where $g$ characterizes the polariton-exciton interaction strength, $\eta$ the energy relaxation \cite{wouters2012energy,berloff2013universality}, \blue{$b$ is proportional the ratio of polariton-reservoir to polariton-polariton interactions and inversely proportional to the reservoir scattering rate}, and where $\gamma$ represents rate of dissipation. The nonresonant pump source is given by pump intensity $P({\bf r},t)$, and the resonant pumping (at $n:1$ resonance with the natural frequency of the cavity) is described by the pumping intensity $\bar{P}({\bf r},t)$ \footnote{Simulations use the fixed parameter values  $g=1$, $b=1$, $\gamma=0.3$, and $\eta=0.3$}. \blue{In these dimensionless units, the healing length is unity while the unit of length is $1\mu m$.}

We begin by focusing on the regime in which the reservoir dynamics react quickly to the condensate wavefunction  ($\partial_tN_R\approx 0)$, in which the energy relaxation $\eta\ll1$, and in which $b\approx 1$. From here we insert $N_R=P/(1+|\psi|^2)$ into Eq. \ref{GPE}; focusing first on the behavior of the system in zero-dimensional space, the dynamics of the polariton condensate are reduced to the following complex ordinary differential equation:

\begin{eqnarray}
 i \dot{\psi} &= &
\frac{(1-i \gamma ) \left| \psi\right| ^2+\left| \psi\right| ^4-i \gamma +(g+i) P}{(\left| \psi\right| ^2+1)/\psi}+ i \bar{P} \psi^{*(n-1)}.
 \end{eqnarray}
Interoducing the Madelung transformation $\psi(t)=\sqrt{\rho(t)}\exp{[i\theta(t)]}$ and separating real and imaginary parts of the resulting equation, we can rewrite Eq.~3 as the real, coupled ODEs

\begin{eqnarray}
\dot{\rho}&= & 
2\frac{\bar{P}\cos (n \theta) \left(\rho ^\frac{n}{2}+\rho ^{\frac{n}{2}+1}\right)+ (P-\gamma )\rho-\gamma  \rho ^2}{(1+\rho)},
 \label{rho}\\
\dot{\theta}&= & \frac{\bar{P}\sin  (n \theta) \left(\rho ^{\frac{n}{2}-1}-\rho ^{\frac{n}{2}}\right)-g P -\rho -\rho^2  }{(1+\rho)}\label{theta}.
 \end{eqnarray}

\begin{figure}[t]
\centering
\includegraphics[width=\columnwidth]{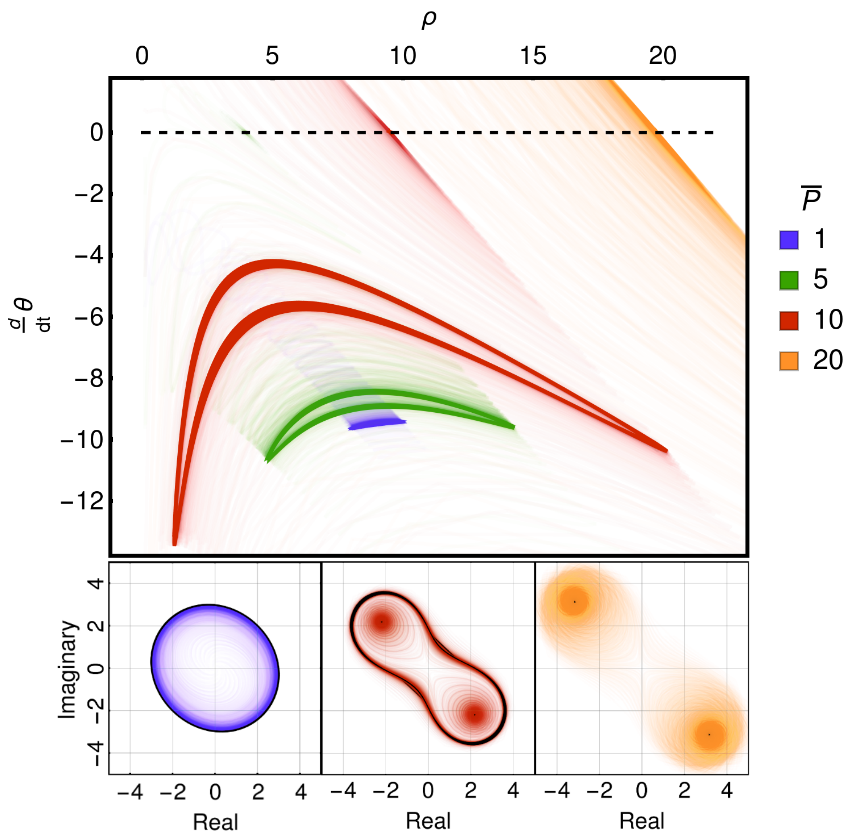}
     \caption{Top: Trajectories of Eqs.(~\ref{rho}-\ref{theta}) traced numerically in phase space from many initial conditions, for fixed parameters $\gamma=1/2$, $P=5$, $g=1$ and various $2^{nd} - $ order resonant pump strengths $\bar{P}=\{1,5,10,20\}$. For clarity, the line of zero phase velocity is marked (dashed black). Bottom: To show the geometry of the symmetry breaking, the same phase trajectories are plotted in the complex plane. From both, we see the transition in behavior from small, nearly uniform oscillations driven by small resonant pumping (blue), to the dual fixed point attractors seen under high resonant forcing (orange). In between, both fixed points and large nonuniform density oscillations are seen (red).}
  \label{phase}
\end{figure}

Physically, $\rho$ represents the time dependent condensate wavefunction density and $\theta$ its phase. Eqs.~(\ref{rho}-\ref{theta}) are not explicitly solvable, but in the limit of small resonant pump strength $\bar{P}$, we can view solutions as perturbations of the steady states familiar to purely nonresonantly pumped condensates. Such nonzero steady states have constant phase evolution $\theta=\mu t$\red{, with frequency $\mu$}. Substituting into Eq.~5 and setting $n=2$ yields the small $\bar{P}$ approximation for the condensate density under simultaneous nonresonant and $2^{nd} - $ order resonant forcing
\begin{eqnarray}
 \rho(t) &= &
\frac{1}{2}\left(\sqrt{\xi(t)^2-4(g P+\xi(t)) }-\xi(t) -1\right)
 \end{eqnarray}
in which $\xi(t)=1+\mu+\bar{P} \sin  (2 \mu  t)$. From here, it is clear that for $\bar{P}=0$, $\xi(t)$ reduces to $\mu+1$ and Eq.~6 returns the familiar steady state solution, with the density fixed by the parameterization. However for $\bar{P}>0$ oscillations take hold, with period given by $T=\pi/\mu$, and with amplitude scaling with $\bar{P}$. Later we will see that even in full 2D simulations of Eqs.~(\ref{GPE}-\ref{NR}), these simple predictions remain robust.

In the other extreme is the scenario of very strong resonant pumping. In this case we expect $n$ fixed points -setting the left hand sides of Eqs.~(\ref{rho}-\ref{theta}) to zero yields $n$ \red{solutions for $\theta$} and one \red{solution} for $\rho$. These are the $n^{th}$ order phase locking solutions. To probe the full range of behaviours beyond these extreme regimes, we numerically integrate Eqs.~(\ref{rho}-\ref{theta}). As this case is of the greatest interest, we again fix $n=2$. Integrating for many initial conditions $\{\rho_i,\theta_i\}$, phase space trajectories are collected for varying resonant pump strength $\bar{P}$ (with other system parameters fixed), shown in Fig.~\ref{phase} (top). The bottom panel of that figure shows some of the same trajectories (matched in color), but represented in the complex plane as opposed to the phase space. In these spaces geometrical interpretations of the effect of resonant pumping, and the resulting density oscillations, become clear. As expected, for $\bar{P}=0$ (not shown), there is a single fixed point attractor corresponding to a single point of nonzero phase velocity and nonzero density in the phase space, which corresponds to a circular trajectory in the complex plane; this is merely the plane wave solution. At the other extreme, high resonant forcing (orange) leads to a set of two fixed point attractors (resolved in the complex plane) at states with fixed density and null phase velocity (resolved in phase space). 
 
The most interesting behaviour is seen between these extremes. As the resonant forcing strength is increased gradually from zero, the smooth \textit{stretching} of the closed state trajectories in the complex plane is observed, in the directions of the symmetry broken fixed points which then eventually form. In this way the geometry of the resonant forcing terms in Eqs.~(\ref{rho}-\ref{theta}) are clear: in the complex plane the terms $\bar{P}\cos{(n\theta)}$ and $\bar{P}\sin{(n\theta)}$ are linear scaling operators, acting along $n$ axes separated by $2\pi/n$. Thus we should expect an $n$-fold stretching of the circular orbit as $\bar{P}$ is increased from zero, and for increasing $\bar{P}$ we should expect the density  increasingly dependent on the phase with degeneracy $n$. Fig.~\ref{phase} confirms these behaviors. For small resonant pumping, we see slight $n$-fold deformation of the orbit in the complex plane (blue), which becomes severely deformed (but with equal symmetry) as the resonant pumping is increased (red). That case also shows the overlap between the limit cycle and phase locked regimes.
We note that the same procedure for any $n^{th}$ order resonance yields the same fundamental result, but showing $n$-fold symmetry in the complex plane. We confirm this in numerical experiments for \red{$n\in \{1,...,5\}$}. 
The warping of the phase-space trajectories has more than a geometric effect: the non-circular closed path in the complex state space is trivially indicative of density oscillations in the wave function. Thus it can be useful to think of the orbit deformations as \textit{driving} the density oscillations (and which fully characterizes their wave-forms). In this way, breathing is a result of the tension between the two natural states of competing symmetries, the U$(1)$ phase symmetry of $\bar{P}=0$, and the $\mathbb{Z}_n$ symmetry of large $\bar{P}$.

\textit{2D Breathers- }So far we have established that under the simultaneous resonant and nonresonant forcing of a polariton condensate, there generically exists a regime of density oscillations in between the plane-wave and phase locked solutions. We now turn to the full, spatially extended (2D) system. The order of resonant pumping, and thus the geometry of the phase-symmetry breaking, is intimately connected with the types of stable topological defects that are allowed in a spatially extended system. In a phase symmetric system, all phases are equally stable, and thus stable topological defects in the form of continuous helical phase gradations wrapped around zero dimensional singularities - this is the celebrated quantum vortex. In a system phase locked by strong $2^{nd} - $ order resonant forcing, there are two equally stable phases differing by $\pi$, so that stable one-dimensional topological defects naturally form between domains of opposite phase (domain walls or `dark solitons'). For $n>2$, more phases become stable, quickly approximating the U$(1)$ symmetry. Thus from the perspective of pattern formation, the $2^{nd}-$order resonant forcing is the most extreme case, as the associated $\mathbb{Z}_2$ symmetry is the starkest departure from U$(1)$ while maintaining the necessary degeneracy. Thus while the our breathing mechanism applies to higher resonances, in the 2D case we will focus only the $2^{nd}-$order resonant forcing.

 \begin{figure}[!t]
\centering
\includegraphics[width=\columnwidth]{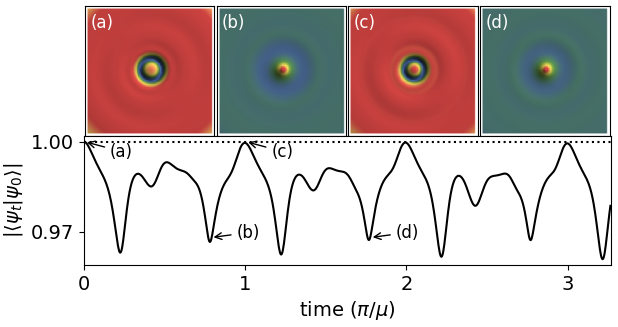}
 \caption{Top: Density of breathing ring soliton at several times, excited by a phase imprinted perturbation in the uniform condensate, forced homogenously with nonresonant and $2^{nd} - $ order resonant pumping. Bottom: overlap between the evolving wavefunction of the condensate and that at the time fixed at (a), showing the periodicity of the complex wavefunction. $P=\bar{P}=5$.}
  \label{rings}
\end{figure}

 \begin{figure}[!b]
\centering
\includegraphics[width=\columnwidth]{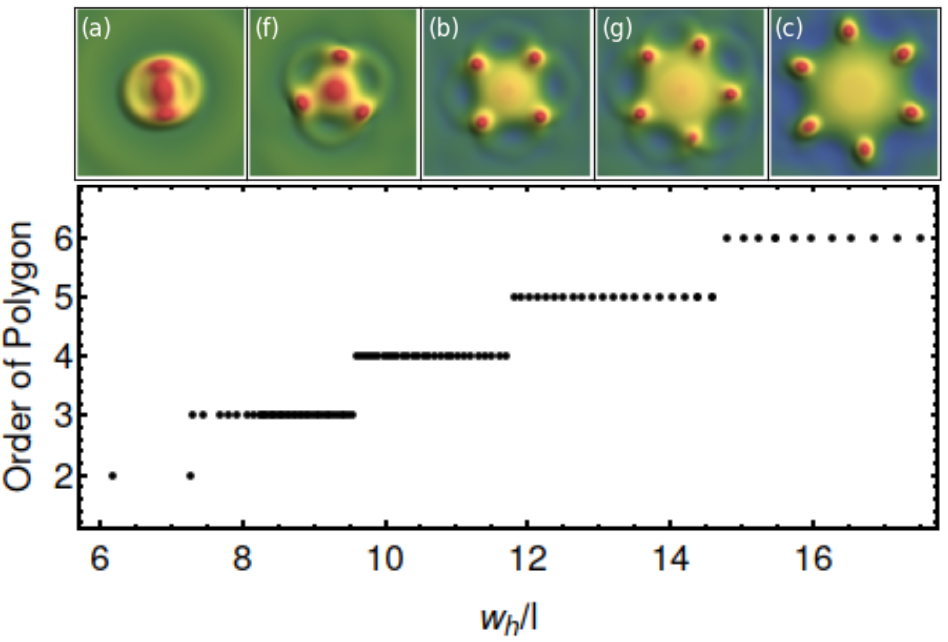}
 \caption{Top: From direct numerical integration of  Eqs.~(\ref{GPE}-\ref{NR}), density profiles exhibiting \red{$m\in\{2,...,6\}$} spatial symmetry, adopted spontaneously for fixed homogenous nonresonant pump ($P=2$) and $2^{nd}-$order resonant pumping with Gaussian profile \red{$\bar{P}\exp[-\alpha r^2]$} at fixed strength ($\bar{P}=15$) and varying \red{half-width} parameter $\alpha$. Bottom: Corresponding dependence of spontaneously adopted symmetry order $m$ on the Gaussian half-width in units of the healing length.}
  \label{depend}
\end{figure}

We begin this by considering the condensate forced uniformly with nonresonant and $2^{nd}-$order resonant (from this point ``resonant") forcing. It was recently shown that ring-shaped breathers can form in such a system \cite{alperin2020formation}, and we now show that these result from the breathing mechanism described in this Letter. In full numerical integration of Eqs.~(\ref{GPE}-\ref{NR}) \footnote{We use fourth-order Runge-Kutta integration, with the physical parameters $\eta=0.3$, $g=1$, and with $\gamma=0.3$}, we prepare a uniform disk-shaped condensate pumped in such a manner, with resonant forcing high enough such that the regime of phase locking is achieved. By phase imprinting a perturbation, we observe the excitation of a breathing ring soliton, as shown in Fig.~\ref{rings}, which have the periodicity $\pi/\mu$ as predicted in the zero-dimensional problem. These structures may thus be interpreted as the localized excitations of the phase-locked state into the limit cycle in a phase space which, as in the zero-dimensional space, admits both simultaneously.

\textit{Polygon Breathers- } One of the powers of polaritonic systems is that pumping can take on any optically feasible profile. We can thus consider the case of spatially dependent resonant forcing, so that the degree to which the phase is symmetry broken can vary spatially.

 \begin{figure}[!t]
\centering
\includegraphics[width=\columnwidth]{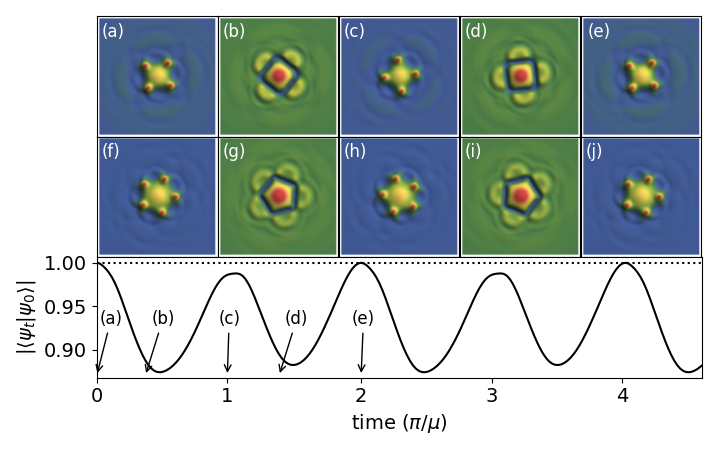}
 \caption{From direct numerical integration of  Eqs.~(\ref{GPE}-\ref{NR}), density profiles over the evolution of two breathing structures exhibiting different degrees of quantized spatial symmetry breaking, under uniform nonresonant pumping $P=2$ and a Gaussian $n=2$ resonant pump of the form $\bar{P}\exp[-\alpha r^2]$, with $\bar{P}=10$. (a)-(e) shows the dynamics of the breather formed when $\alpha=0.05$, which spontaneously adopts degree-4 polygonal symmetry, and then evolves in a closed cycle: the bottom panel shows the inner product of the condensate wavefunction over time with that shown in (a), showing that the wavefunction perfectly repeats periodically. (f)-(j) show the evolution of a degree-5 symmetric breather, formed spontaneously when $\alpha=0.03$.}
  \label{dense}
\end{figure}

We thus consider the scenario of a large disk-shaped region of uniform nonresonant pumping, with a resonant pump of Gaussian profile $\bar{P}\exp[-\alpha r^2]$ at the centre of that region, with $\alpha$ characterizing the inverse width of the pump, so that the degree of the symmetry breaking of the phase depends on the radial distance from the center of the pump. This is the most extreme when $\bar{P}$ and $\alpha$ are chosen such that the condensate wavefunction is forced into the phase locked regime at the centre, but can be seen to transition into the regime in which the symmetry breaking is negligible. With direct numerical integration of Eqs.~(\ref{GPE}-\ref{NR}), we simulate this geometry. For large $\alpha$ (small spot), density oscillations are driven around the center at the radius at which the condensate is in the breathing regime, forming a breathing ring. 

For larger resonant pump spots however, the behavior changes drastically: as the pump spot width is increased (keeping $\bar{P}$ constant), the radius of the dark ring increases, and existing out of the bistable regime, reaches the circumference at which the ring becomes unstable to the ``snake-instability", the well known phenomenon in which azimuthal modes of the annular defect shatter the dark soliton into an integer number of chiral defects \cite{theocharis2003ring,carr2006vortices}. This instability thus naturally quantizes the number of vortex-antivortex pairs produced as a function of the ring radius. This quantization is demonstrated in Fig.~\ref{depend}, which shows the dependence of the emergent polygonal symmetry as a function of the Gaussian \red{half-width at half maximum (half-width)} in units of healing lengths, as determined from numerical experiments. \red{The half-width is defined as $w_h = \sqrt{\log(2)/\alpha}$. }.

 \begin{figure}[!b]
\centering
\includegraphics[width=\columnwidth]{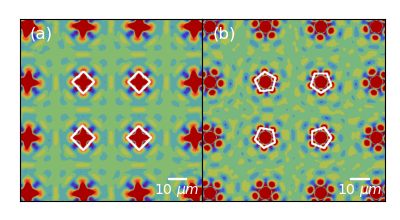}
 \caption{Detail of condensate under uniform nonresonant pumping, simultaneously pumped by a square lattice of resonant pumps with Gaussian profile $\bar{P}\exp[-\alpha r^2]$. Orientations of central lattice site excitations highlighted in white. (a) Spontaneous orientational order emerges when $\alpha$ is chosen such that the lattice site excitations match the symmetry of the lattice. (b) A glassy ordering emerges instead when $\alpha$ is chosen such that the symmetry of the lattice site excitations is incommensurate with that of the lattice. $P=1.5$ and $\bar{P}=10$. $\alpha=0.1$ in (a) and $\alpha=0.075$ for (b). Lattice length is set at $35\mu m$, with periodic boundaries.}
  \label{crystal}
\end{figure}

Once the rotational symmetry breaks after a finite number of oscillation cycles, the symmetry remains broken, and a new dynamically stable structure is formed which, though evolving dynamically, is at every time symmetric under transformations of the dihedral group D$_m$ (the group of symmetry transformations of the polygon of degree $m$). Fig. \ref{dense} shows density profiles of two polygonal breathers at several times during their evolutions. The bottom panel of that figure shows the inner product \red{$|\braket{\psi_0|\psi_t}|=\frac{1}{A}|\int \psi^*(\boldsymbol{r},0) \psi(\boldsymbol{r},t) d\boldsymbol{r}|$ (where A is chosen such that $|\braket{\psi_0|\psi_0}|=1$)} of the $m=4$ (square-symmetric) breather, where we arbitrarily set $\psi_{0}$ to the wavefunction at the time shown in (a). This shows that the wavefunction indeed forms a closed periodic cycle once the symmetry has broken, despite the rotational symmetry of the physical system. We note that the periodicity of the breather is almost exactly equal to twice the predicted periodicity of the density oscillations studied in 0D. \red{ This period doubling comes from the broken rotational symmetry of the structure: we observe that $ \ket{\psi(t) }=R(\pi/m) \ket{ \psi(t+\pi/\mu)}$ where the operator $R(\phi)$ rotates the breather by $\phi$ radians about its center, so that $\ket{\psi(t) }=\ket{ \psi(t+2\pi/\mu)}$.}

\textit{Orientation Glass- } While these spontaneously polygon-symmetric excitations are translationally fixed by the location of a Gaussian resonant pump, they do posses a rotational degree of freedom. As breathers, the polygonal structures radiate density oscillations through the condensate, and these radiation patterns possess the polygonal symmetry of their source. We might thus imagine the emergence of orientational order in a lattice of polygonal breathers. Fig.~\ref{crystal} shows the direct numerical simulation of a condensate forced by uniform nonresonant pumping and by a square lattice of Gaussian pumps with \red{inverse} width parameter $\alpha$. When $\alpha$ is chosen such that the lattice sites exhibit square-symmetric excitations, the symmetries of the excitations and the lattice are commensurate and spontaneously align (a). When $\alpha$ is instead chosen such that the lattice sites exhibit pentagonal symmetry, the symmetries of the lattice and the lattice sites are incommensurate, causing geometric frustration resulting in a glassy state. These are the analogues of the ferromagnetic and spin glass states, but where the spin degree of freedom (parameterized by $\mathbb{Z}_n$ for discrete spins) is replaced by the polygonal orientation degree of freedom (parameterized by $D_m$).

In conclusion, we have introduced a generic mechanism of breather formation in nonequilibirum condensates forced by simultaneous resonant and nonresonant pumping. In the case of $2^{nd}$-order resonant pumping, we have shown that this mechanism can lead to highly nontrivial dynamical behaviour, including the spontaneous adoption of unusual spatial symmetries and emergent order. We hope that our Letter sparks interest in the physics of condensates forced by multiple driving frequencies, and that it sheds some light on the importance of the underlying symmetries imposed by those forces, especially on emergent spatial symmetries.

\bibliography{alperin}
\end{document}